\documentclass[twocolumn,amsmath,nofootinbib,superscriptaddress]{revtex4}

\usepackage{url}
\usepackage{amssymb}
\usepackage{amsfonts}
\usepackage{amsbsy}
\usepackage{graphicx}
\usepackage{hyperref}
\usepackage{array}
\usepackage{multirow}
\usepackage{epsfig}
\usepackage{graphics}
\usepackage{amssymb}
\usepackage{amsmath}
\usepackage{comment}
\usepackage{enumerate}
\usepackage{subcaption}
\usepackage{natbib}
\usepackage{xcolor}
\usepackage{fancyhdr}
\def\spose#1{\hbox to 0pt{#1\hss}}
\def\simlt{\mathrel{\spose{\lower 3pt\hbox{$\mathchar"218$}}
   \raise 2.0pt\hbox{$\mathchar"13C$}}}
\def\simgt{\mathrel{\spose{\lower 3pt\hbox{$\mathchar"218$}}
     \raise 2.0pt\hbox{$\mathchar"13E$}}}
 \def\simpropto{\mathrel{\spose{\lower 3pt\hbox{$\mathchar"218$}}
     \raise 2.0pt\hbox{$\propto$}}}

\def\beq#1{\begin{equation}\label{#1}}
\def\eeq{\end{equation}}
\def\beqa#1{\begin{eqnarray}\label{#1}}
\def\eeqa{\end{eqnarray}}

\fancypagestyle{plain}{
\fancyhf{}
\fancyfoot[C]{\fontfamily{cmss}\selectfont UNCLASSIFIED}
\fancyhead[C]{\fontfamily{cmss}\selectfont UNCLASSIFIED}
\fancyhead[R]{\fontfamily{cmss}\selectfont
  \fcolorbox{black}{white}{\parbox{1.9in}{\raggedleft\small
    \color{black}DISTRIBUTION STATEMENT A.\\Approved for public release.}}}
}

\parskip=\medskipamount

\begin{document}
\title{Network Traffic Anomaly Detection Using Recurrent Neural Networks}
\author{Benjamin J. Radford}
\email[Corresponding author: ]{benjamin.radford@keywcorp.com}
\affiliation{Sotera Defense Solutions, a KeyW Company}
\thanks{This research was developed with funding from the Defense \mbox{Advanced} Research Projects Agency (DARPA). The views, opinions and/or findings expressed are those of the authors and should not be interpreted as representing the official views or policies of the Department of Defense or the U.S. Government.}
\author{Leonardo M. Apolonio}
\affiliation{Cybraics}
\author{Antonio J. Trias}
\affiliation{DZYNE Technologies}
\author{Jim A. Simpson}
\affiliation{Cynnovative}

\begin{abstract}

We show that a recurrent neural network is able to learn a model to represent sequences of communications between computers on a network and can be used to identify outlier network traffic. Defending computer networks is a challenging problem and is typically addressed by manually identifying known malicious actor behavior and then specifying rules to recognize such behavior in network communications. However, these rule-based approaches often generalize poorly and identify only those patterns that are already known to researchers. An alternative approach that does not rely on known malicious behavior patterns can potentially also detect previously unseen patterns. We tokenize and compress netflow into sequences of ``words'' that form ``sentences'' representative of a conversation between computers. These sentences are then used to generate a model that learns the semantic and syntactic grammar of the newly generated language. We use Long-Short-Term Memory (LSTM) cell Recurrent Neural Networks (RNN) to capture the complex relationships and nuances of this language. The language model is then used predict the communications between two IPs and the prediction error is used as a measurement of how typical or atyptical the observed communication are. By learning a model that is specific to each network, yet generalized to typical computer-to-computer traffic within and outside the network, a language model is able to identify sequences of network activity that are outliers with respect to the model. We demonstrate positive unsupervised attack identification performance (AUC 0.84) on the ISCX IDS dataset which contains seven days of network activity with normal traffic and four distinct attack patterns.

\end{abstract}

\maketitle

\thispagestyle{plain}

\date{\today}
\vspace{10mm}

\maketitle

\section{Introduction}

Defending computer networks from unauthorized use has become an increasingly critical challenge for governments, industries, and private individuals in recent years. An estimated \$75 billion were spent globally on cybersecurity services and solutions in 2016 \citep{idc:2016}. Meanwhile, the threat actors targeting vulnerable systems have expanded from the perceived lone-wolf hackers of the 1990's and early 2000's to include advanced and well-funded criminal groups and state actors. Previous intrusion detection system (IDS) paradigms that relied on signature-based matching against known attacks and attack vectors are no longer sufficient. An alternative approach to cybersecurity detection frames the problem as one of anomaly detection; a model of network activity is estimated and future activity on the network is evaluated with respect to its probability under the learned model. We use a long short-term memory (LSTM) recurrent neural network (RNN) to learn ordered sequences of network traffic representative of a computer network and then evaluate the ability of this model to detect malicious activity on that network. We demonstrate that LSTM RNNs are able to detect patterns of traffic indicative of malicious computer system use without the assistance of labeled training data and without visibility into each machine's internal state or processes. Further, we provide evidence that unsupervised models can detect traffic indicative of malicious activity even when they are trained on network data that are not representative of a pristine (attack-free) traffic. This should boost confidence in the use of machine learning for cybersecurity applications, a domain for which pristine training data are costly and rarely available.
\section{Background}

We motivate our research by drawing on cutting-edge research in two active fields: cybersecurity and natural language processing (NLP). Cybersecurity is a broad field and different networks pose different challenges to researchers and practicioners. The challenge of interest to us is briefly detailed alongside alternative perspectives on cybersecurity. We also note that cybersecurity applications have, in recent years, seen an increase in the use of machine learning and statistical models to accomplish tasks that were previously often performed via signature-based matching methods.  Building on this trend, we note similarities between the cybersecurity problem as represented by network logs (i.e. netflow) and the challenges inherent to modeling natural languages. 

\subsection{Cybersecurity}

Cybersecurity encompasses a wide range of problems that includes, but is not limited to, intrusion detection, malware detection, preventative security (e.g. access controls, 2 factor authentication, security training), network monitoring, and associated investigative and remediation efforts. While we believe that threats manifest in a number of ways and may be invisible to any single detection method or even combination of methods, we nonetheless focus our efforts here solely on network monitoring to identify anomalous network traffic. 

Network traffic data generally consist of logs summarizing the communications between network-connected devices. Often these data are aggregated such that the available information includes a start time and duration for the communications represented by a single record. Each record includes two Internet Protocol addresses (IP addresses) representative of network-connected machines. Due to the dynamic assignment of IP addresses and to the actual structure of networks themselves, it cannot be assumed that any given IP maps consistently to the same physical device or even to any single device at a given time. Additionally, network logs generally include measures of bytes and packets transferred during the duration of a communication as well as the ports and protocols used in the communication. Network logs of this type are frequently collected via dedicated hardware installed at strategic points within the network of interest.

Because network traffic logs (flow) can be collected passively by dedicated hardware, their collection is typically invisible to the users of other hardware on that network. Network logs are also relatively small when compared to other log types available for cybersecurity purposes like host-based logs or full packet capture. Combined, the unobtrusive nature of these data combined with their (relatively) small storage footprint make them a popular choice for organizations implementing a cybersecurity posture. 

Because of their ubiquity, a number of tools take advantage of network traffic logs for cybersecurity monitoring. Signature-based tools can utilize network traffic logs to validate traffic against blacklists of known-bad IPs, to monitor the volume of traffic and enforce volume or rate-based limits, and to monitor port and protocol usage to identify the use or attempted use of services that should not be present.  Somewhat more complex approaches include the use of SQL-like queries to perform operations on flow data to make rule-based matching more precise or customized to a particular network's needs. For instance, regular expressions might be used to parse user agent strings or internet addresses to identify keywords or tokens of interest. 

Recent work has used machine learning to circumvent the need for rule sets that are necessarily specified a priori. 

Machine learning has also been applied to other cybersecurity-relevant data types. For example, Veeramachaneni et al. present a system for user-in-the-loop machine learning on web logs and firewall logs \citep{veeramachaneni:etal:2016}.  Researchers have demonstrated that LSTM RNNs can be trained on system processes and utilized for intrusion detection \citep{kim:etal:2016}. Supervised network traffic and attack classification has also been demonstrated with LSTM RNNs.\citep{kim:etal:2016b}.

\subsection{Natural Language Processing}

Neural network models have recently achieved state-of-the-art performance on a number of NLP tasks. The popular \emph{2vec} family of models including word2vec and doc2vec, among others, has been applied to great success in projecting sparse and high-dimensional natural language representations into low-dimensional continuous-valued vector spaces while retaining the semantic and syntactic relationships of the original language \citep{mikolov:etal:2013a, le:mikolov:2014}. Another line of research has approached language modeling as a sequence problem in the vein of Markov models and suffix tree language models \citep{shannon:1948, kennington:etal:2012}. Recent work in this area has utilized LSTM RNNs for language modeling \citep{sundermeyer:ney:2015}. Combining LSTM RNNs with character-level convolutional neural networks (CNNs) has been shown to produce results comparable to the state-of-the-art with fewer parameters than comparable models \citep{jozefowicz:etal:2016}. 

Network flow metadata share characteristics of natural language. Communications between networked devices are captured in ordered sequences, and we expect these communications to follow a set of rules, similar to a grammar, determined by the services and protocols they utilize. However, the underlying grammatical rules are typically obscured from the analyst and so explicitly modeling them is impossible. Therefore, unsupervised language models are a natural choice for inferring the data generating processes of network metadata.

\section{Data}

We utilize a public dataset for intrusion detection (IDS) tasks from the University of New Brunswick's Canadian Institute for Cyberseurity (CIC) and the Information Security Centre of Excellence (ISCX), hereafter referred to as ISCX IDS \citep{shiravi:etal:2012}. The dataset represents seven days of simulated network traffic with a variety of attack behaviors including infiltration, denial of service (DoS), distributed denial of service (DDoS) via IRC Botnet, and brute force SSH attacks. The raw data, provided in the form of full packet capture (PCAP) is roughly 90 gigabytes. 

We are interested in anomaly detection via network flow metadata and so focus on a pre-processed flow-style dataset that accompanies the full ISCX IDS data. This metadata table contains just over 2 million flow records occurring between June 11, 2010 and June 17, 2010. After de-duplicating some records, we are left with 1.9 million entries. These metadata consume only 412 megabytes of disk space and make clear the storage cost (not to mention computational cost) advantage of flow over PCAP. 

From these flow data, we produce ordered sequences of flows per IP-pair (dyad). Dyads are undirected and so the pair $\text{IP}_a \text{IP}_b$ is equivalent to $\text{IP}_b \text{IP}_a$. Within each sequence, a single flow record corresponds to a single token, a word in the NLP analogy. We perform two sets of analyses that require distinct sequences. In the first method, tokens are of the form \texttt{Protocol:$floor(log_2(bytes))$}. An example sequence may therefore look like \texttt{IP$_a$IP$_b$: TCP:10|TCP:12|UDP:04}. We refer to these as proto-byte sequences. In the second set of analyses, we look to ports as our tokens. In particular, we use heuristics to determine which port, per flow, is likely to be the service port. For each port pair, we retain the lowest port value and drop the higher port value. We also collapse all ports above 10,000 to a single token as a rough approximation of ephemeral or otherwise uncommon ports. A sequence of ports for a given IP-pair looks like \texttt{IP$_a$IP$_b$: 80|80|443|80}. There are 168,218 dyad hours in the data.

Sequences are formed with a rolling window applied only within (not across) dyad-hours. A dyad-hour consists of all flow records between two IP addresses within a single hour. Aggregating to the dyad-hour allows us to make high-resolution predictions of attacks; we recognize that dyads may exhibit malicious behavior during some time periods and non-malicious behavior during others and hope to be able to distinguish between them. The beginning of each sequence is zero-padded and a mask is applied at the modeling stage to avoid biasing results due to the padding. Dyad-hour units are classified as attacks if they contain at least one record labeled ``attack'' in the source data.\footnote{An alternative is to label those dyad-hour units as attacks if the sum of ``attack'' records is greater than the sum of ``normal'' records. In practice, we find that there are very few such IP-pair-hours in the data and therefore opt for the simpler aggregation rule.} Our models are fully unsupervised with respect to ``attack'' labels; we use these labels only for validation and not for model training. 
\section{Methodology}

\begin{figure}
\includegraphics[width=\linewidth]{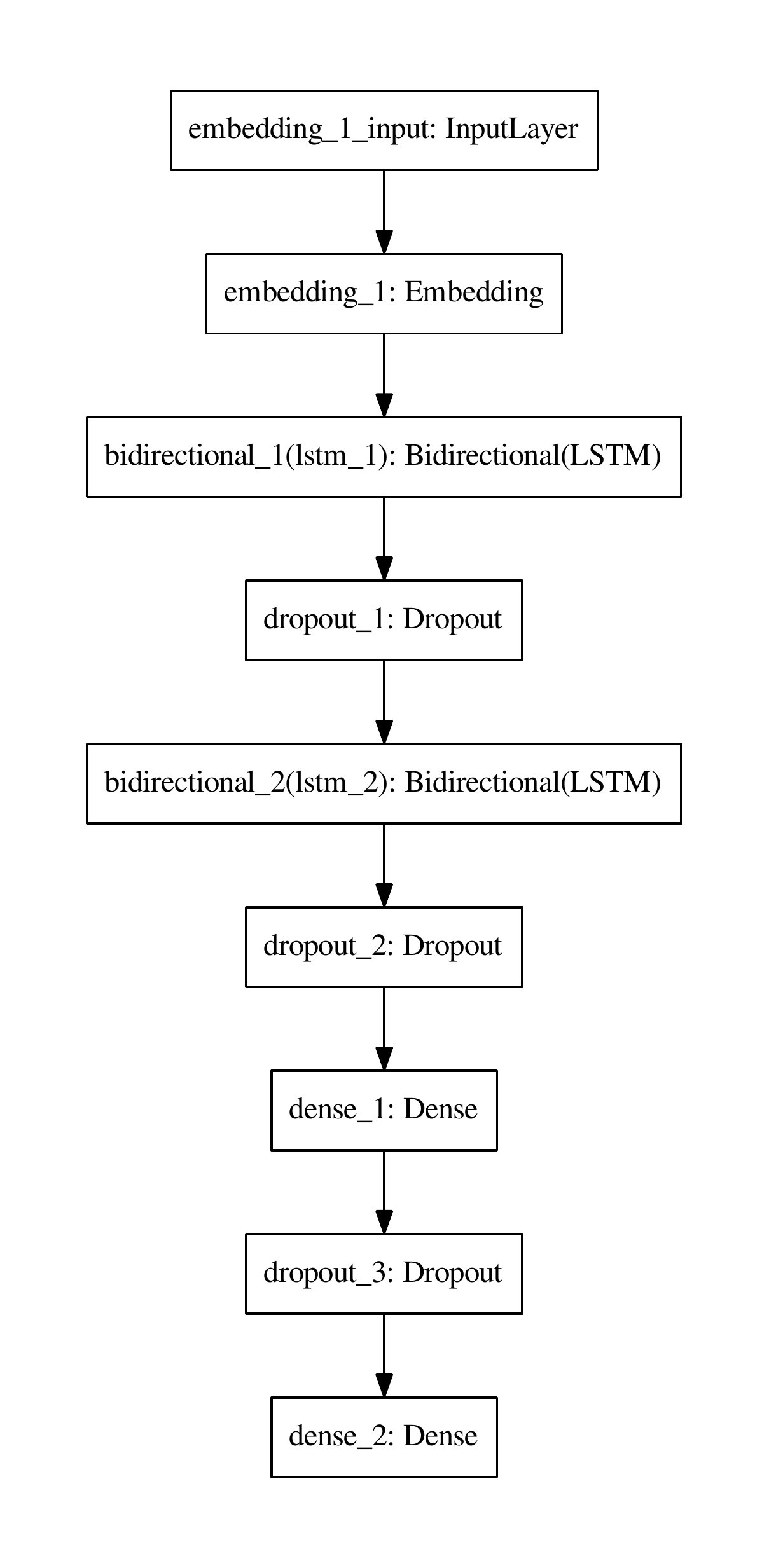}
\caption{Model architecture.}\label{fig:model}
\end{figure}

We use a LSTM RNN to model flow sequences. Our model comprises two stacked, bidirectional, LSTM layers, a single dense layer activation, and a single fully-connected softmax output layer. A 20\% dropout rate is applied between each layer. Each LSTM layer is composed of 50 hidden cells with linear activation and on the first layer and rectified linear activation on the second layer. In addition, an initial embedding layer projects input sequences from $V$ unique tokens into a dense 100-dimensional vector space. We train on the entire dataset with a ten-token sliding window. For each ten-token window, the model is trained to predict the subsequent (eleventh) token. Left-censored sequences are zero-padded and zero-masking is applied in the training stage to prevent biasing the model. The model is diagrammed in Figure~\ref{fig:model}. 

Unsupervised cybersecurity applications require learning a model of the typical network behavior against which newly-observed behaviors can be evaluated. Presumably, malicious activities will manifest as anomalies with respect to the baseline model. For some networks, it may be conceivable that a ``clean'' baseline can be made; it may be possible to collect data on the network during a period for which it is likely that no attempts at network misuse or abuse have occurred. However, for many networks, ensuring a clean collection period for baseline model estimation may be infeasible or impossible. We test and compare both scenarios here.

In the \emph{clean baseline} test, we train our network model on the first day and a half of the ISCX dataset for which we know that no attacks are present. In the \emph{dirty baseline} test, we train our network on the full ISCX dataset such that attack and non-attack traffic are both learned by the model. Once a model of the network traffic is learned, we use that model to predict values for every flow in the dataset. We use multiclass logarithmic loss as an ``outlier score;'' we assume that poorly-predicted datapoints are more likely to be those associated with anomalous or malicious traffic. The maximum observed logarithmic loss value per IP-dyad hour is taken to be that observation's outlier score.\footnote{Our choice to use the maximum observed value is motivated by both design considerations and observed performance. Using the maximum observed outlier score per dyad-hour to score that dyad-hour matches our decision to consider dyad-hours as ``attacks'' if they contian at least a single flow record that is designated ``attack.'' In other words, our outlier score aggregation method corresponds to our validation data aggregation method. Additionally, we observed that the maximum value aggregation rule performed better with respect to our selected model performance metrics than did the averaging of ourlier scores. For a discussion of aggregation rules and outlier ensembles, see \cite{zimek:etal:2014}.}

In addition to the \emph{clean} and \emph{dirty baseline} tests, we also conducts tests in which DoS and DDoS attacks have been omitted from the data. Because flows are not labeled by attack category in the data, we opt to remove the entirety of both June 14 and June 15 from the dataset to ensure that the DoS and DDoS attacks are not present for training or model evaluation. We refer to these tests as \emph{NoDoS}. It is possible that DoS and DDoS attacks, by virtue of their high-volume traffic profiles, could result in the models overfitting to these types of attacks and therefore being unable to flag them as anomalies. Alternatively, if the models can accurately flag DoS and DDoS attacks, the volume of these activities will inflate performance metrics due to the imbalance of DoS-like attack flows versus other attack type flows. 
\begin{figure*}
\begin{subfigure}{.5\textwidth}
\includegraphics[width=\linewidth]{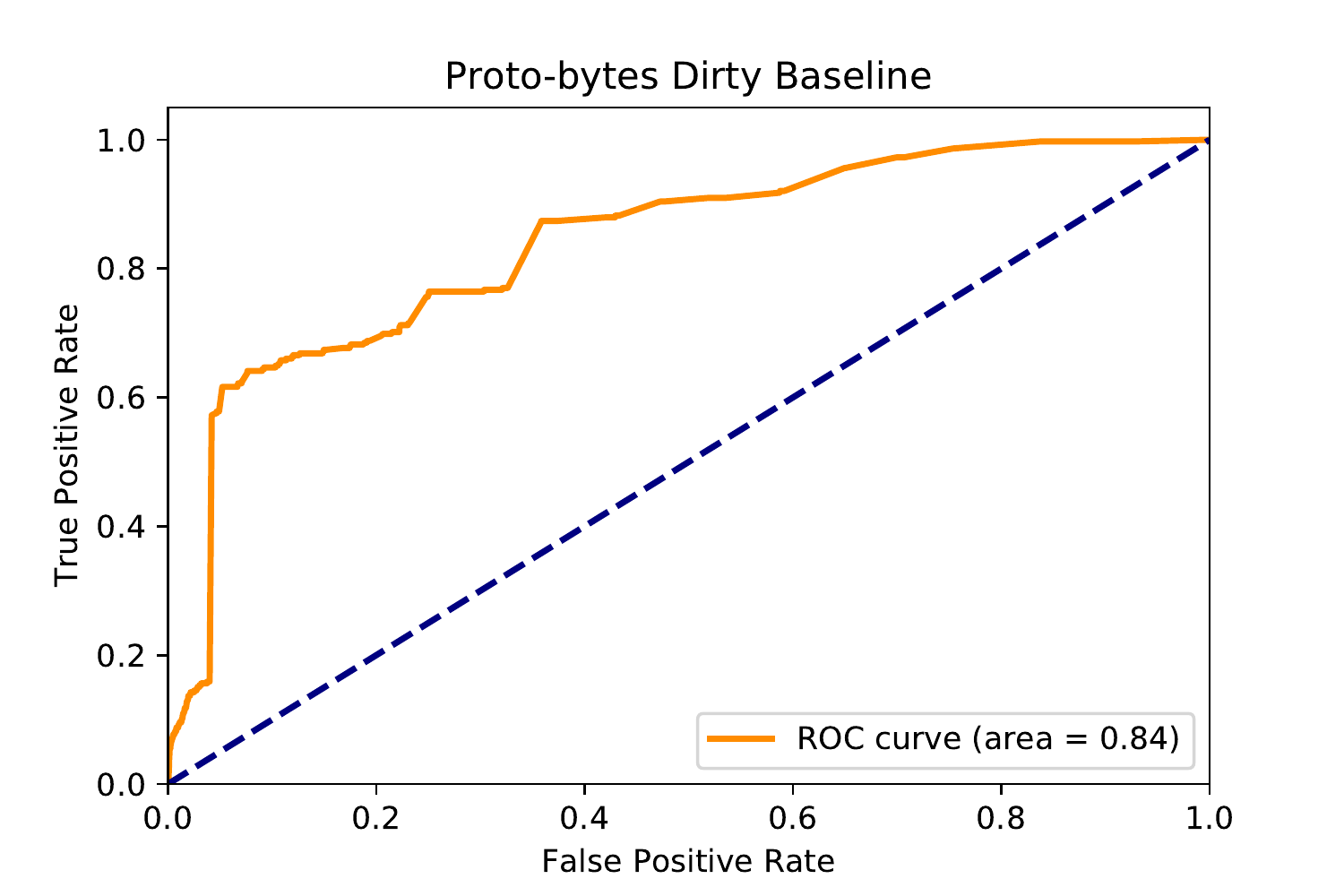}
\caption{Proto-byte features.}
\end{subfigure}%
\begin{subfigure}{.5\textwidth}
\includegraphics[width=\linewidth]{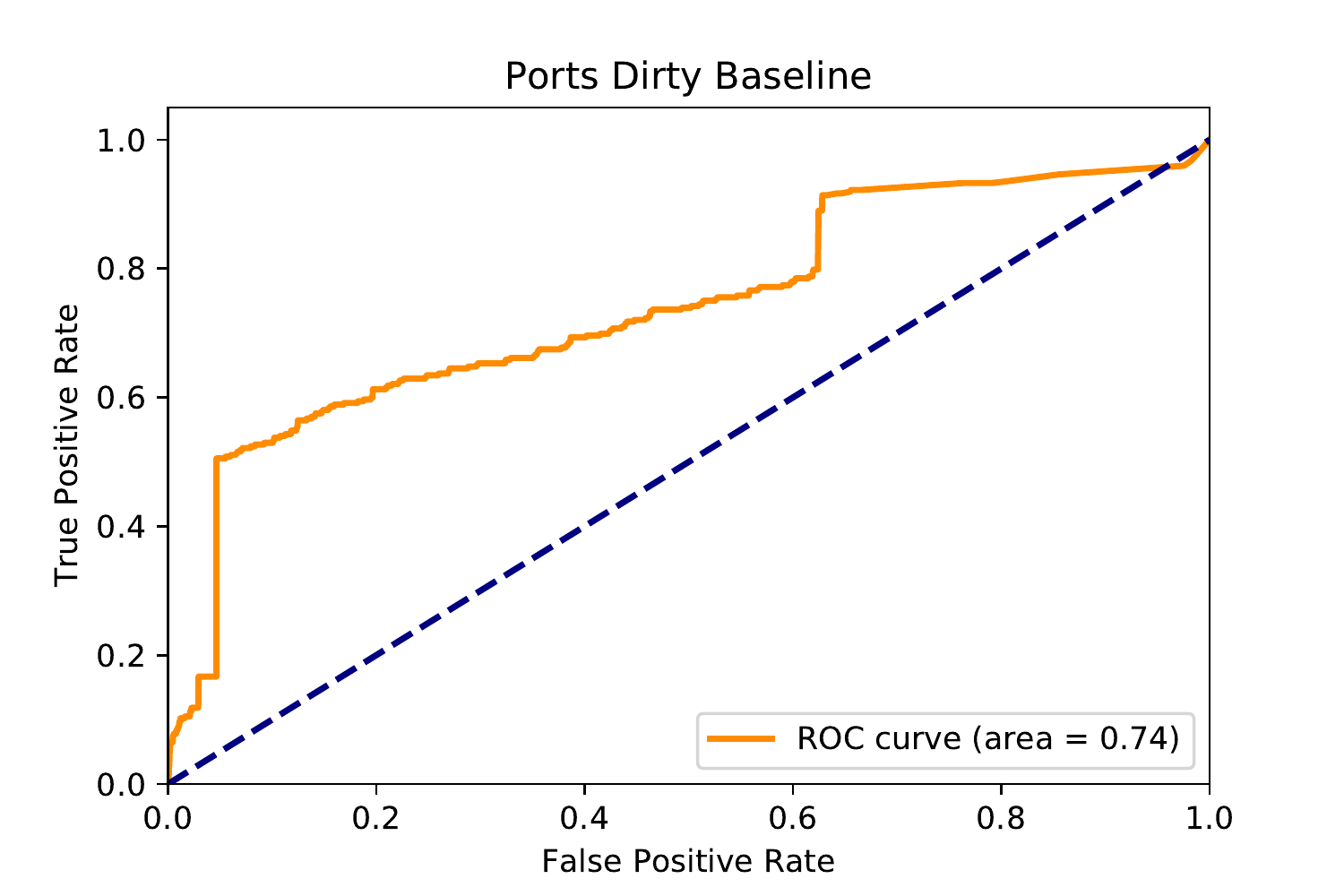}
\caption{Service port features.}
\end{subfigure}\caption{ROC plots for \emph{dirty baseline} models.}\label{fig:roc_dirty}
\end{figure*}

\begin{figure*}
\begin{subfigure}{.5\textwidth}
\includegraphics[width=\linewidth]{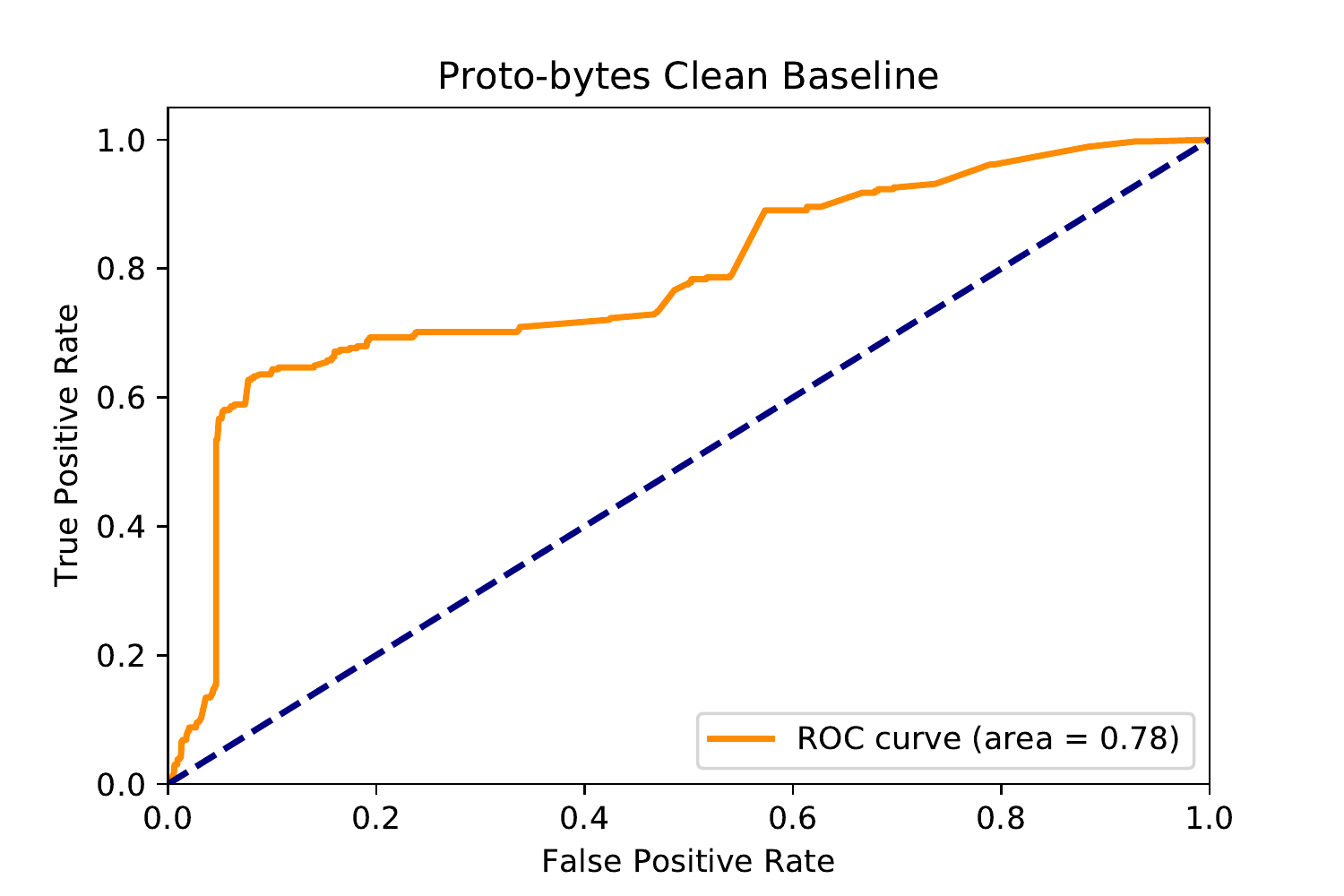}
\caption{Proto-byte features.}
\end{subfigure}%
\begin{subfigure}{.5\textwidth}
\includegraphics[width=\linewidth]{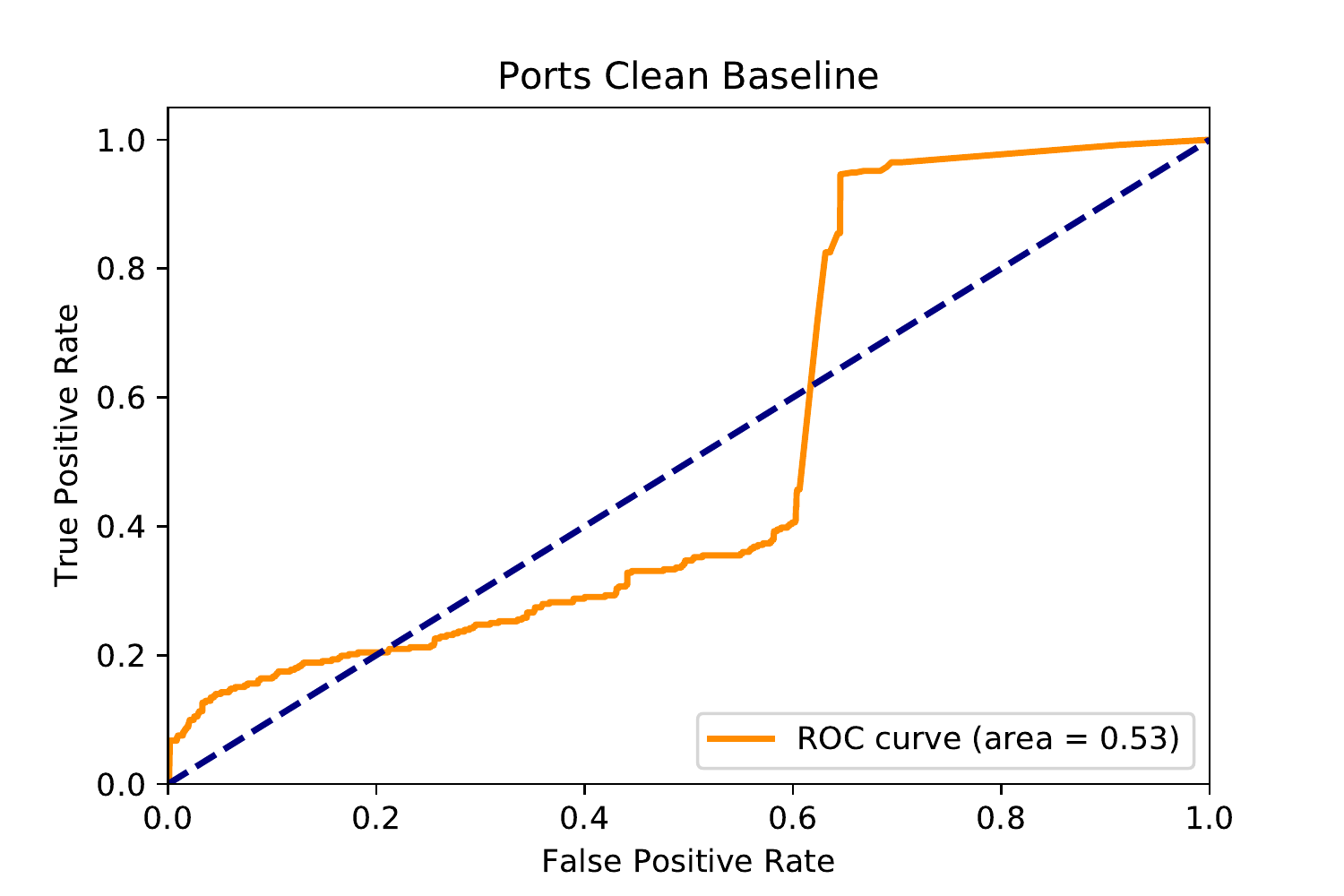}
\caption{Service port features.}
\end{subfigure}\caption{ROC plots for \emph{clean baseline} models.}\label{fig:roc_clean}
\end{figure*}

\begin{figure*}
\begin{subfigure}{.5\textwidth}
\includegraphics[width=\linewidth]{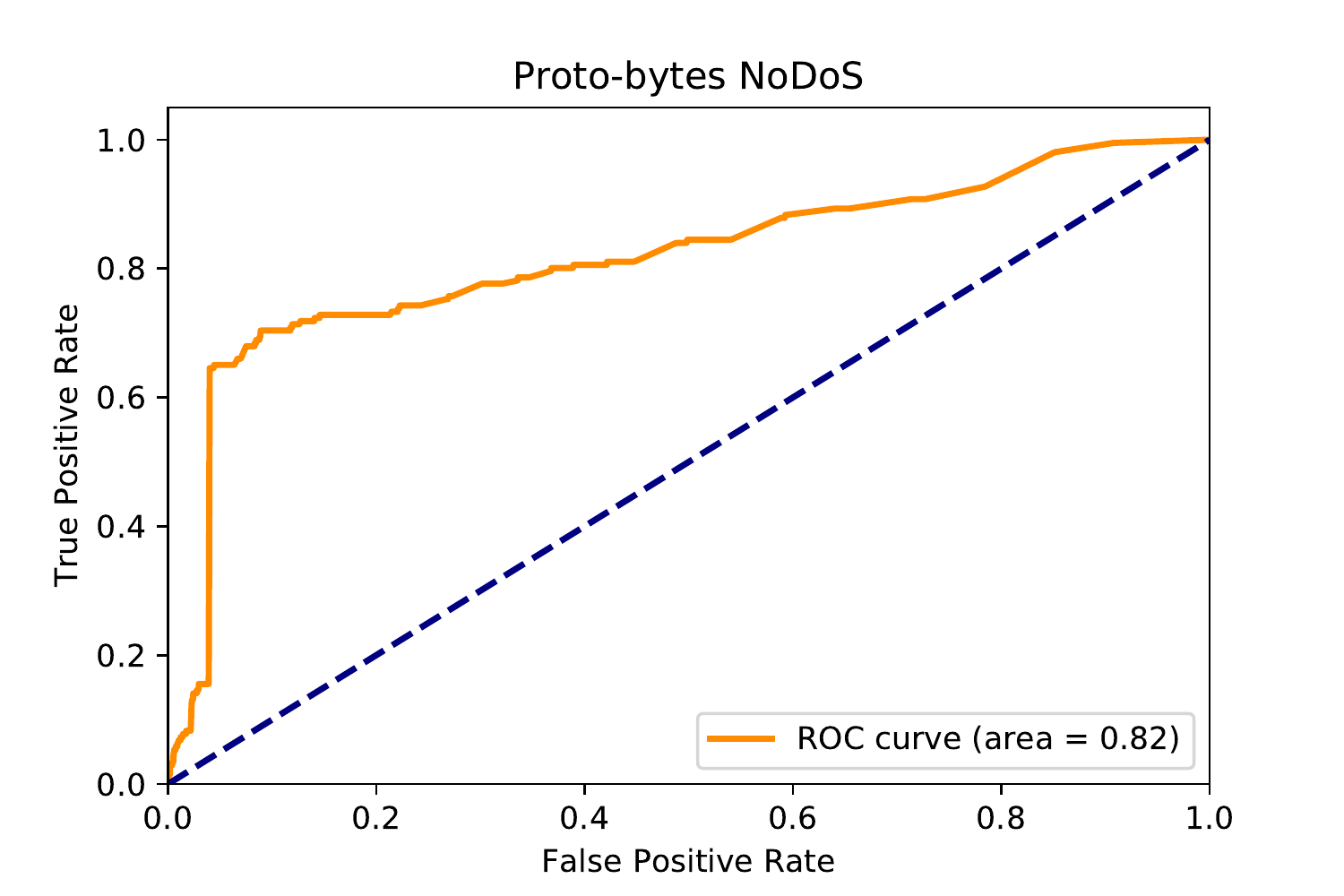}
\caption{Proto-byte features.}
\end{subfigure}%
\begin{subfigure}{.5\textwidth}
\includegraphics[width=\linewidth]{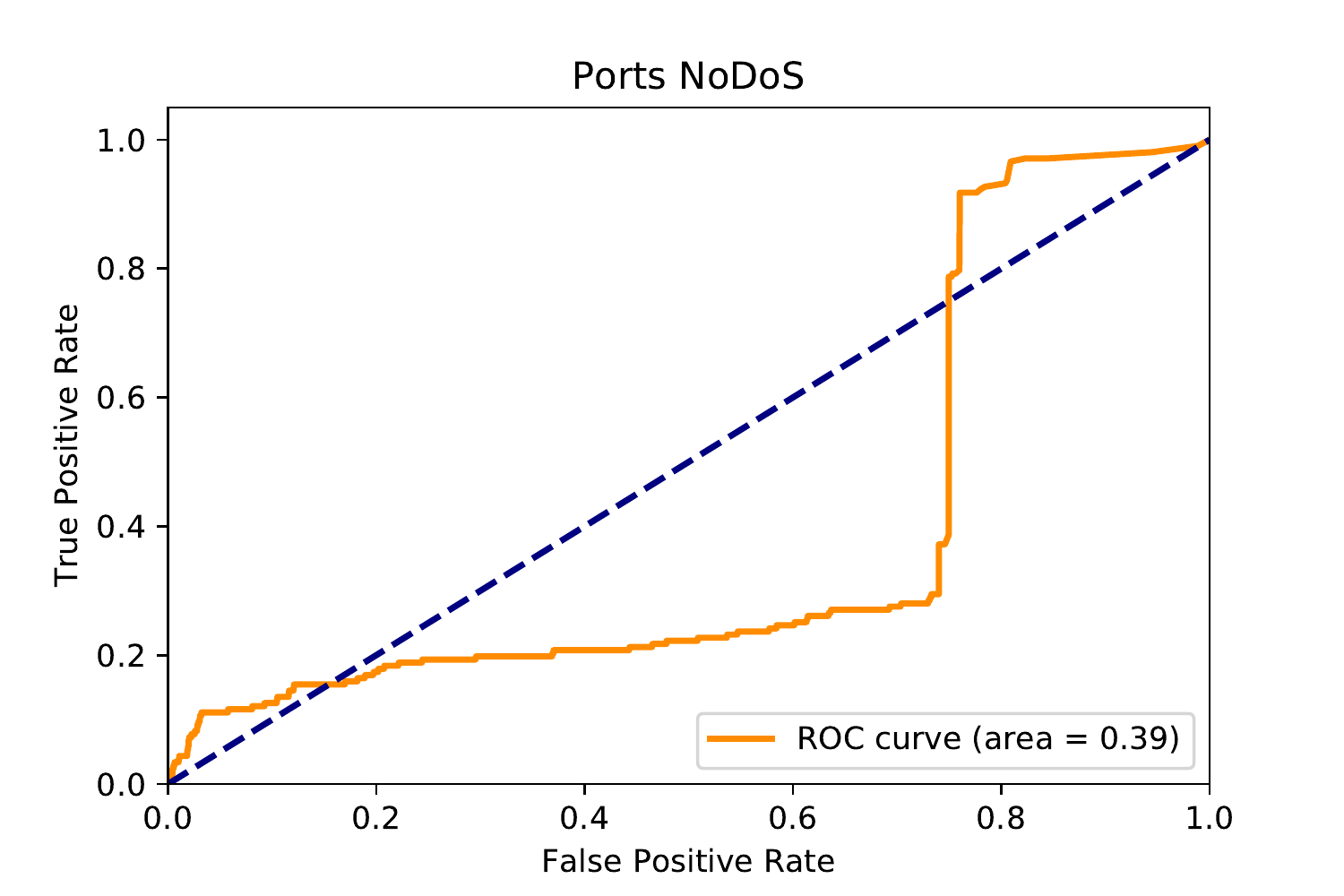}
\caption{Service port features.}
\end{subfigure}\caption{ROC plots for \emph{NoDoS} models.}\label{fig:roc_noddos}
\end{figure*}

\section{Results}

Our models are able to identify anomalous network traffic in both the \emph{clean baseline} and \emph{dirty baseline} scenarios.  Model performance is depicted with Receiver Operating Characteristic (ROC) plots in Figures~\ref{fig:roc_dirty} through \ref{fig:roc_noddos}. ROC plots depict the False Positive Rate (FPR) versus the True Positive Rate (TPR) over all possible threshold values. If the observed curve in a ROC plot falls along the diagonal line $x=y$, the model is assessed to perform no better than random chance. A perfect curve is one that forms a right angle at FPR=0.0 and TPR=1.0 which would indicate that all assigned probabilities to the positive class (here malicious activity) are greater than assigned probabilities to the negative class (here benign activity). The Area Under the Curve (AUC) is the value attained by integrating over the ROC curve. An AUC of 1.0 indicates perfect classification and an AUC of 0.5 indicates random chance. Models that perform better than chance should attain AUC values in the range $\left(0.5, 1.0\right]$. 

We elect to use AUC as our evaluation metric because we want to respect the unsupervised nature of our stated problem. Because we do not have labeled data with which to calibrate a threshold and we offer no unsupervised method for selecting an anomaly or attack threshold, we prefer a performance metric that is agnostic to threshold. Without an understanding of the baseline rate of attack observations in the raw data, for example, we cannot estimate an appropriate threshold with which to hard partition the data and calculate accuracy. The outlier scores produced by our method, logarithmic loss, can be thought of as an ordered list that ranges from non-anomalous to anomalous $\left[ 0.0, + \infty \right)$. 

We find that the protocol-bytes feature set outperforms the service port feature set in all tested scenarios. Looking at subfigure a of Figures~\ref{fig:roc_dirty} through \ref{fig:roc_noddos} we can see that all models based on proto-byte sequences produce higher AUC scores than any model based on service port sequences. This indicates that dyad-hours that contain flows labeled ``attack'' are, on average, scored higher with respect to outlier score than ``non-attack'' dyad-hours. This holds even for the \emph{NoDoS} model in which DoS and DDoS attacks have been removed. This is encouraging because it indicates that proto-byte sequences can identify attack types that are not inherently high-traffic and high-byte-count.

The relatively poor performance of service port sequences in identifying cyberattack activity is puzzling. The relatively high AUC score of the \emph{dirty baseline} service port model indicates that this feature set does in fact distinguish normal network behavior from attack behaviors. Why the performance of models based on service port sequences drops off in the subsequent models, \emph{clean baseline} and \emph{NoDoS}, is unclear. 

We also find that the \emph{dirty baseline} models outperform \emph{clean baseline} models in all cases. This should be encouraging for cybersecurity practitioners who may be unable to obtain a sample of guaranteed clean traffic from their network. It is indicative that well-trained models are able to identify, at least in some circumstances, malicious network traffic even when those behaviors are present in the model's training data. We suspect that the \emph{dirty baseline} models may outperform the \emph{clean baseline} models due to training dataset size. The selected deep LSTM models require the estimation of large number of parameters and model performance may suffer when the sample dataset is insufficiently large to precisely estimate those parameters. Because the \emph{dirty baseline} training dataset is much larger than the \emph{clean baseline}, those models may better learn the common behaviors of the network. In subsequent work, we hope to use subsampling of the \emph{dirty baseline} training data to test this hypothesis. 

The \emph{NoDoS} proto-byte sequence model performs substantively the same as the \emph{dirty baseline} proto-byte sequence model with respect to attack detection. This provides some reassurance that proto-byte sequences, as a feature set, are robust to attack type or characteristic. 

While we evaluate overall model performance here, we also recognize that positive performance results do not necessarily guarantee efficacy in field applications. High AUC scores, for instance, may indicate that most attacks in the data are scored as such by the model, but that high false positive rates nonetheless make the inspection of these results tedious, time-consuming, or impossible.
\section{Conclusion}

We have demonstrated that network behaviors can be learned from traffic metadata using LSTM RNNs and applied for anomaly detection (i.e. cybersecurity) purposes. This work is important because it offers cybersecurity practitioners an effective and unsupervised tool for network protection that requires the collection and storage of only readily-available and relatively cheap network metadata. Furthermore, our method learns a model of the actual network to be protected and we believe it is therefore widely applicable across a variety of computer network infrastructures and architectures. 

In future research we hope to validate our method in real-world settings against known adversaries. The most effective model presented here (\emph{dirty baseline} proto-byte sequences) is trained on the full dataset to be scored, a method that we recognize will not translate well into an operational setting. Instead, we recommend future researchers in this area consider a streaming or mini-batch framework in which a model is learned on a temporal subset of network data and used for scoring new data as it is generated. This should minimize computational requirements as training the LSTM can be done periodically and when compute resources are cheap or otherwise available. We also believe that combining unsupervised models with user-in-the-loop feedback mechanisms and supervised learning could provide valuable performance improvements with respect to fewer false positive alerts and greater confidence in the identification of common attack vectors. In addition, we recognize that not all cybersecurity anomalies of interest will necessarily manifest in sequences of flow metadata. Ongoing efforts to conceptualize the network-modeling problem as one of signal processing, clustering, and network analysis have shown promise in identifying various classes of anomalous network behaviors. 

We hope that our efforts here will motivate the necessity for and encourage further research into unsupervised anomaly detection on computer network metadata. Cybersecurity is an open problem and one that is very costly to businesses, institutions, governments, and individuals, all of whom rely daily on the integrity of networked systems. We believe that incorporating unsupervised network modeling and monitoring techniques with existing signature-based cybersecurity solutions will enable cybersecurity practitioners to finally get ahead of malicious actors by improving our ability to detect adversary behavior in a timely manner even when adversaries utilize previously unknown vulnerabilities. 

\bibliographystyle{IEEEtran}
\bibliography{master}
\end{document}